\documentclass[elsarticle,preprintnumbers,amsmath,amssymb,floatfix,nofootinbib]{revtex4}
\usepackage{epsf}
\usepackage{graphicx}
\usepackage{subfigure}
\usepackage{color}
\begin{document}
\title{Detecting a heavy neutrino electric dipole moment at the LHC}
\author{Marc Sher and Justin R. Stevens}

\affiliation{Physics Department, College of William and Mary, Williamsburg VA  23187}

\date{\today}

\begin{abstract}
The milliQan Collaboration has proposed to search for millicharged particles by looking for very weakly ionizing tracks in a detector installed in a cavern near the CMS experiment at the LHC.    We note that another form of exotica can also yield weakly ionizing tracks.  If a heavy neutrino has an electric dipole moment (EDM), then the milliQan experiment may be sensitive to it as well.  In particular, writing the general dimension-5 operator for an EDM with a scale of a TeV and a one-loop factor, one finds a potential EDM as high as a few times $10^{-17}$~e-cm, and models exist where it is an order of magnitude higher.   Redoing the Bethe calculation of ionization energy loss for an EDM, it is found that the milliQan detector is sensitive to EDMs as small as $10^{-17}$ e-cm.     Using the production cross-section and analyzing the acceptance of the milliQan detector, we find the expected $95\%$ exclusion and $3\sigma$ sensitivity over the range of neutrino masses from $5-1000$~GeV for integrated luminosities of $300$ and $3000\ {\rm fb}^{-1}$ at the LHC.
\end{abstract}
\maketitle

\section{Introduction}
The overwhelming evidence for dark matter shows that physics beyond the Standard Model must exist, and yet the LHC has yet to find evidence for this new physics.    One alternative which has been increasing in popularity is the possibility of a ``dark sector" in which there is a sector that does not directly interact with Standard Model particles (see Section 6.22 of Ref.~\cite{Hewett:2012ns} and the extensive list of references therein).   In such a sector, there can be one or more $U(1)$ groups, and the dark photon will, in general, mix with the conventional hypercharge gauge boson.  This mixing, $\epsilon$, is generally small, and results in particles obtaining very small charges of $O(\epsilon)$ times the usual electron charge.

There have been numerous searches for  dark photons~\cite{Hewett:2012ns, Merkel:2011ze,Abrahamyan:2011gv,Corliss:2017tms,DeNapoli:2017aps}  and astrophysical and cosmological bounds on millicharged particles have been studied~\cite{Davidson:1991si,Davidson:2000hf,Mohapatra:1990vq,Davidson:1993sj,Dubovsky:2003yn,Dolgov:2013una,Vogel:2013raa}.    Additionally, some experimental searches for millicharged particles have been done~\cite{Davidson:2000hf, Prinz:1998ua,Diamond:2013oda,Perl:2009zz}.   However, these searches focus on the sub-GeV mass region.    Recently, a new experiment dedicated to searching for millicharged particles of much higher masses has been proposed at the LHC~\cite{Haas:2014dda,Izaguirre:2015eya,Ball:2016zrp}.  The milliQan experiment will consist of layers of scintillator detector situated in the Observation and Drainage gallery above the CMS experimental cavern.  It is designed to search for very weakly ionizing tracks as expected from millicharged particles.   Details are found in the Letter of Intent~\cite{Ball:2016zrp}.    It will be sensitive to charges as low as 0.3\% of the electron charge, over the mass range from 100 MeV to 100 GeV, a range that is currently unexplored.
Another experiment near the LHCb detector, MoEDAL,~\cite{Mitsou:2017doz}
 is currently taking data, and while it is sensitive to millicharged particles, the luminosity is much smaller than that of milliQan.

Theorists have proposed a large number of unusual exotica that can be searched for at the LHC, including magnetic monopoles ~\cite{Mitsou:2017doz,Katre:2016jid}, black holes~\cite{Park:2012fe}, long-lived charged particles~\cite{Hamaguchi:2006vu,Asai:2009ka,Chen:2009gu,Huitu:2010uc}, etc.   An attractive feature of the MoEDAL experiment is that it is sensitive~\cite{Mitsou:2017doz} to a wide variety of exotica.    The milliQan experiment is much more focused and is designed only to search for millicharged particles.     The purpose of this paper is to point out that the milliQan experiment will also be extremely  sensitive to another form of exotica:  the possible electric dipole moment (EDM) of a heavy neutrino.

The possibility that a heavy neutrino could have a large EDM was discussed fifteen years ago in Ref.~\cite{Sher:2002ij}.   It was noted that several models have leptonic EDMs scaling as the cube of the mass, and an explicit model was exhibited with a neutrino EDM of $O(10^{-16})$ e-cm.     The fact that such a large EDM could occur is not surprising.    Writing the effective low-energy dimension-five operator as
\begin{equation}
\frac{c}{\Lambda}\bar{\nu}_L\sigma_{\mu\nu}i\gamma_5N_RF_{\mu\nu}
\end{equation} 
then if $\Lambda = 1$ TeV and $c$ is $O(1)$, one finds an EDM of approximately $10^{-15}$ e-cm.    In a realistic model, one expects this to be suppressed by at least one loop, but large EDMs are certainly not impossible.    By ``neutrino", we refer to a heavy neutral Dirac fermion, without regard for whether it is an isosinglet or isodoublet.

In Ref.~\cite{Sher:2002ij}, it was noted that there are few phenomenological bounds on such an EDM, especially if the neutrino is vectorlike.   At the time, there was great interest in a high energy linear $e^+e^-$ collider.    Studies of detection through $e^+e^-\rightarrow \bar{\nu}\nu\gamma$ were carried out, and the production cross section in $e^+e^-$ colliders was calculated, assuming the neutrino was in a vector-like isodoublet.    It was also noted there that such a neutrino would leave a weak ionization track in a detector.    However, no studies from a hadron collider were presented, and there were, at the time, no experimental searches sensitive to a large EDM.

In Section~\ref{sec:ionloss}, the calculation of the ionization loss of a neutrino with an EDM is presented, and it is seen that an experiment like milliQan could be sensitive, in principle, to EDMs as low as $10^{-17}$ e-cm.   In Section~\ref{sec:production}, the cross-section for neutrino pair production (primarily via Drell-Yan production through a virtual photon) is calculated and used to estimate the sensitivity of the milliQan experiment.  Finally, Section~\ref{sec:conclusion} contains our conclusions.

\section{Ionization loss}
\label{sec:ionloss}

Except through missing energy-momentum, detection of long-lived neutral weakly interacting particles at the LHC is impossible.    However, a neutrino with a large  EDM does interact electromagnetically and thus can lose energy in a detector.   This ionization loss was discussed in Ref.~\cite{Sher:2002ij} and we will follow their argument closely, with some minor modifications to account for relativistic effects.    The derivation of the Bethe formula for ionization loss is given clearly in Jackson~\cite{Jackson:1998nia}.   This formula is, of course, derived for a Coulomb interaction and we generalize it to an EDM.

Suppose a heavy neutrino travels in the $x$-direction and an atomic electron is at an impact parameter $y=b$.     The impulse given to the electron will depend on the orientation of the dipole.   In practice, one should consider a dipole in an arbitrary direction, however we will look at each of the three directions and average appropriately.   Suppose the dipole is oriented in the $z$-direction, perpendicular to the plane of the neutrino motion and the electron.    The electric field component (for distances greater than the size of the dipole moment, which is the case here) is only in the $z$-direction, and the impulse given to the electron is $\Delta\vec{p}=\int_{-\infty}^{+\infty} e\vec{E}\ dt$  with 
\begin{equation}
E_z = \frac{e D}{4\pi\epsilon_0}(b^2+v^2t^2)^{-3/2},
\end{equation}   
\noindent where $t=0$ is the time of closest approach and $eD$ is the size of the EDM.   Integrating gives an impulse of $\frac{e D}{4\pi\epsilon_0}\frac{2}{vb^2}$.      Now suppose the dipole is in the $y$-direction.    The electric field components are now
\begin{equation}
E_x = \frac{e D}{4\pi\epsilon_0r^3}(3\sin\theta\cos\theta) \qquad E_y = \frac{e D}{4\pi\epsilon_0r^3}(3\cos^2\theta-2)
\end{equation}
\noindent where $r^2=b^2+v^2t^2$ and $\cos\theta\equiv \frac{b}{r}$.  Integrating gives an impulse in the $x$-direction which vanishes (as expected by symmetry) and the impulse in the $y$-direction which is also $\frac{e D}{4\pi\epsilon_0}\frac{2}{vb^2}$.    Finally, if the dipole is in the $x$-direction, the impulse vanishes.

Thus, the impulse if the dipole is in the plane perpendicular to the neutrino's motion is $\frac{e D}{4\pi\epsilon_0}\frac{2}{vb^2}$ and the impulse vanishes if the direction is parallel to the neutrino's motion.   For a large number of interactions, which will be the case here, one thus expects a net average impulse to an electron to be half of this result, giving an impulse of $\frac{e D}{4\pi\epsilon_0}\frac{1}{vb^2}$.     Since the electron is moving non-relativistically, the impulse is converted into an energy transfer 
\begin{equation}
\Delta E =\frac{ |\Delta\vec{p}|^2}{2m} = \frac{e^4D^2}{2m(4\pi\epsilon_0)^2(vb^2)^2}.
\end{equation}
\noindent This must now be cylindrically integrated over the impact parameter.   The maximum energy transfer is $\Delta E = 2m\gamma^2v^2$~\cite{Jackson:1998nia}, thus $b_{min}^2 = e^2D/(2m\gamma v^2 (4\pi\epsilon_0))$ and
\begin{equation}
\frac{dE}{dx} = 2\pi NZ\int_{b_{min}}^\infty \ \Delta E(b) b db = \pi N Z\left(\frac{e^2}{4\pi\epsilon_0}\right) D\gamma.
\end{equation}
\noindent where $N$ is the neutron number and $Z$ is the nuclear charge.
    The usual logarithm in the Bethe formula is absent and the electron mass and neutrino velocity drop out.    Plugging in numbers, this becomes $2.7\times 10^{11}~(D\gamma (N/N_A))~{\rm MeV\ cm}$, where $N_A$ is Avogadro's number.    In the usual units of ${\rm MeV\ {g}^{-1}\ cm^2}$, this becomes $2.7 \times 10^{11}~(D\gamma (Z/A))~{\rm MeV\ {g}^{-1}\ cm^2}$, where $D$ is in units of cm and $A=Z+N$.
    
 What is the discovery potential of the milliQan experiment?   They are sensitive to millicharged particles with charges of roughly $0.003$ times the electron charge, corresponding to an ionization loss of $10^{-5}$ times that of a muon (whose ionization energy loss is about $2~{\rm MeV\ {g}^{-1}\ cm^2}$).      Plugging this in, milliQan could potentially set an upper limit on $D\gamma$ of $8\times 10^{-17}$ cm.     Since $\gamma$, for a neutrino mass of tens of GeV, can be $O(10-100)$, this shows that EDMs in the range of $10^{-17}$ e-cm are certainly accessible.    It is, of course, essential that a reasonable number of these neutrinos be produced, and so we now turn to the production cross-section.
 
 \section{Production and Sensitivity}
 \label{sec:production}
 
 As first noted in Ref.~\cite{Escribano:1996wp}, the relevant operator consistent with gauge invariance involves coupling to the $B_{\mu\nu}$ gauge boson, which contains a $Z$ and a photon.    Given our definition of the EDM, the coupling to the $Z$ will be that EDM times $\tan\theta_W$.   This coupling will have very little effect on the numerical results found here.   In general, there could be an operator coupling the heavy neutrino to the $SU(2)$ field tensor, but this will only occur if the neutrino is in a vector-like isodoublet.   Such neutrinos would have to be above $45$ GeV to avoid large contributions to the $Z$ width, and also would be accompanied by a heavy charged lepton.   Since such an interaction would have an arbitrary parameter, we will only look at the isosinglet case; if the neutrino is in an isodoublet, then we are assuming the coupling is not large enough to affect our results. 
 
 The parton-level cross section for neutrino production will then be through a virtual Drell-Yan photon, $\bar{q}q\rightarrow \gamma^* \rightarrow \bar{\nu}\nu$, where the last vertex occurs through the EDM.   On dimensional grounds, the cross section must be proportional to $D^2$, which already has units of area, and thus one expects the cross section to be constant at high energy.   The total differential cross section is given by
 \begin{equation}
 \frac{d\sigma(\hat{s})}{d\Omega}= \frac{Q_q^2 \alpha^2 D^2}{4} \sin^2\theta \left(1 + \frac{4M_\nu^2}{\hat{s}}\right) \sqrt{1-\frac{4M_\nu^2}{\hat{s}}}
 \label{eqn:xsec}
 \end{equation}
where $M_\nu$ is the heavy neutrino mass,  $Q_q$ is the quark charge in units of the electron charge and $\hat{s}$ is the partonic center-of-mass energy.     Note that if the neutrino were an isodoublet, there would also be a tree-level contribution, independent of the EDM, from a virtual $Z$, and this would increase the cross section substantially.    Note also that the fact that the cross section does not fall as $1/\hat{s}$ means that for very large EDMs, unitarity will be violated.    As noted in Ref.~\cite{Sher:2002ij}, the effective coupling is $\alpha D\sqrt{s}$ and for EDMs below $10^{-15}$ e-cm, this is less than unity, and thus breakdown of unitarity will not be a serious issue.    Another way to see this is that the larger the EDM, the smaller the scale of the new physics which generated the effective operator, and for an EDM larger than $10^{-15}$ e-cm, this scale is smaller than $\sqrt{\hat{s}}$.
 	
The angular distribution varies as $\sin^2\theta$, which differs from that of millicharged particles which have a typical distribution of $1+\cos^2\theta$, and this will provide a method of providing a way to distinguish the possibilities.   Alas, the MoEDAL experiment is forward peaked and is thus less likely to see a neutrino EDM, but the milliQan experiment is at a $45$ degree angle.

To estimate the sensitivity of the milliQan experiment to a heavy neutrino EDM we calculate the leading order proton-proton differential cross section $\frac{d\sigma}{d\tau dy}$, where $\tau \equiv x_1x_2$ and $y\equiv \frac{1}{2}\ln \frac{x_1}{x_2}$.  The partonic cross section in Eqn.~\ref{eqn:xsec} is convolved in the usual way with the parton distribution functions of Ref.~\cite{Gao:2013xoa}, and summed over the $u,d$ and $s$ quarks.  Signal Monte-Carlo events are then generated according to this cross section for a range of neutrino masses ($5-1000$~GeV) and EDMs ($10^{-17}-10^{-15}$~e-cm).  The expected number of heavy neutrino pair events are then simulated for the 14 TeV center-of-mass energy collisions at the high luminosity LHC, assuming an integrated luminosity of $\mathcal{L} = 300$ or $3000~\textrm{fb}^{-1}$.   

\begin{figure}[t]
\begin{center}
\includegraphics[width=0.7\textwidth]{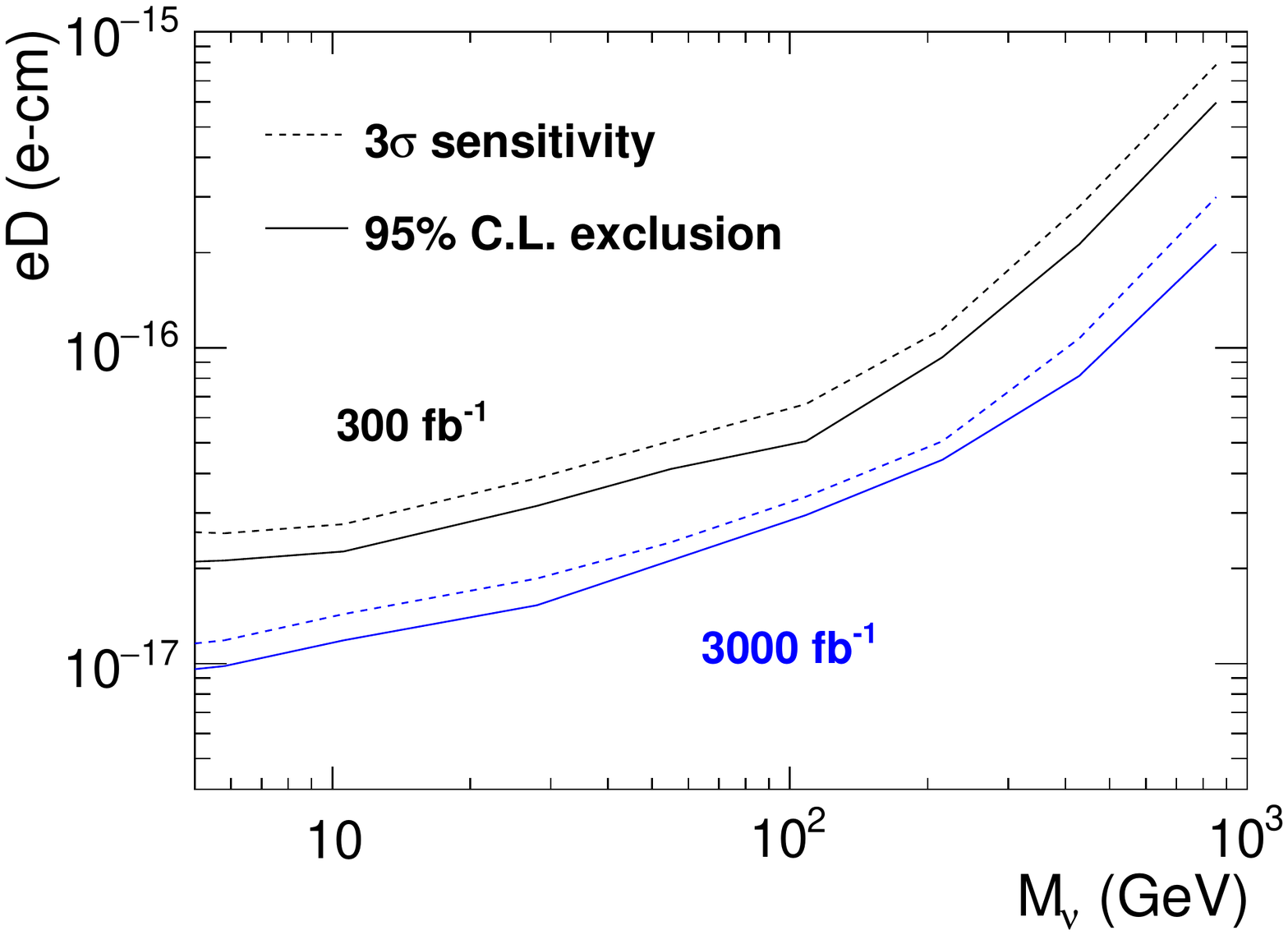}
\end{center}
 \caption {The expected 95\% C.L. exclusion (solid) and $3\sigma$ sensitivity (dashed) for heavy neutrino EDM detection using the milliQan experimental setup at $\sqrt{s}=14$~TeV, assuming $\mathcal{L}=300~(3000)~\textrm{fb}^{-1}$ integrated luminosity in black (blue).} \label{fig:limit}
\end{figure}

The acceptance of the milliQan detector is estimated by requiring the signal events have a heavy neutrino which impinges on the $1~\textrm{m}\times1~\textrm{m}\times3~\textrm{m}$ scintillator array of the experiment at the expected location, described in Ref.~\cite{Ball:2016zrp}.  The expected number of detected signal events is given by the number of heavy neutrinos with $D\gamma > 8\times10^{-17}$~cm, which is consistent with the requirement for millicharge detection sensitivity for the milliQan detector.  The expected background rates are taken from the estimates in the milliQan Letter of Intent~\cite{Ball:2016zrp} as 165 (330) background events for $300~(3000)~\textrm{fb}^{-1}$.  Based on these estimates, in Fig.~\ref{fig:limit} we show the estimated 95\% confidence level exclusion and $3\sigma$ sensitivity of the milliQan experiment to a heavy neutrino EDM. 

We emphasize that the production cross section assumed that the neutrino was a weak isosinglet.   If it is an isodoublet, the production cross section could be much higher, and that could lead to substantial tighter bounds.  Thus, the expected sensitivity displayed in Fig.~\ref{fig:limit} are conservative upper bounds.
 
\section{Conclusion}
\label{sec:conclusion}

Occasionally, major discoveries in physics are made by detectors designed for something completely different - the classic example is the discovery of supernova neutrinos in detectors designed to search for proton decay.   The milliQan experiment is designed to search for millicharged particles, which occur in several appealing models of beyond the Standard Model physics.  Here, we point out that if a heavy neutral fermion has an electric dipole moment, then the same experiment may be sensitive to such particles, and have estimated the sensitivity attainable.

Our sensitivity estimates are based on the expected milliQan detector design parameters~\cite{Ball:2016zrp}.  A more detailed study can be performed by the experimenters to identify if there are additional optimizations that may improve the sensitivity to a neutrino EDM.  From the theoretical point of view, one can improve on the Jackson-level calculation of the ionization loss by considering shell corrections, density corrections and higher order corrections.   This analysis is currently underway.  In addition, the sensitivity will be substantially greater if the neutrino is an isodoublet  (this would, of course, set a lower bound of around $45$ GeV on the mass).

Should milliQan see a signal, of course, one would immediately want to distinguish between millicharged particles and a neutrino EDM.    This would require measuring the angular distribution, which would be difficult for milliQan since it is fixed in a cavern.   There might be an energy dependence that can be studied.    Nonetheless, detection of a positive signal would rapidly lead to new experiments and new detectors.

\section*{Acknowledgements}

This work is supported by the National Science Foundation grant PHY-1519644 and the Department of Energy Early Career Award contract DE-SC0018224.

\end{document}